%% file: omegah.tex
\documentclass[aps,prl,twocolumn,superscriptaddress,showpacs,byrevtex]{revtex4}
\usepackage{epsfig}

\def\DESepsf(#1 width #2){\epsfxsize=#2 \epsfbox{#1}}
\begin{document}

\pagestyle{empty}


\title{\boldmath
Observation of  $B^{\pm} \to \omega K^{\pm}$ Decay
}

\date{\today}

\normalsize

\input{author}

\normalsize

\begin{abstract}
We report the first observation of
the charmless two-body mode $B^{\pm} \to \omega K^{\pm}$ decay,
and a new measurement of the branching fraction
for the $B^{\pm} \to \omega \pi^{\pm}$ decay.
The measured branching fractions are
${\cal B} ( B^{\pm} \to \omega K^{\pm} ) =
    (9.2{}^{+2.6}_{-2.3}\pm 1.0) \times 10^{-6}$ and
${\cal B} ( B^{\pm} \to \omega \pi^{\pm} ) =
    (4.2{}^{+2.0}_{-1.8}\pm 0.5) \times 10^{-6}$.
We also measure the partial rate asymmetry of
$B^{\pm}\to\omega K^{\pm}$ decays and obtain
${\cal A}_{CP} = -0.21 \pm 0.28 \pm 0.03$.
The results are based on a data sample of 29.4 fb$^{-1}$
collected on the $\Upsilon(4S)$ resonance by the Belle
detector at the KEKB $e^{+} e^{-}$ collider.
\end{abstract}




\maketitle

\pagestyle{plain}

\begin{figure}[b]
\begin{center}
\epsfig{file=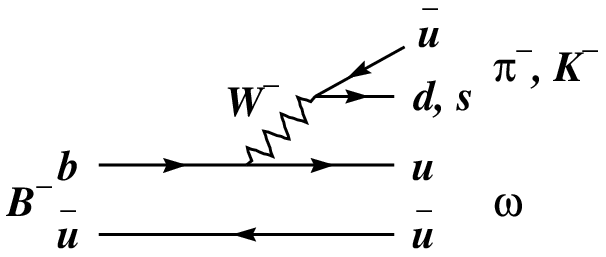, width=1.65in}
\epsfig{file=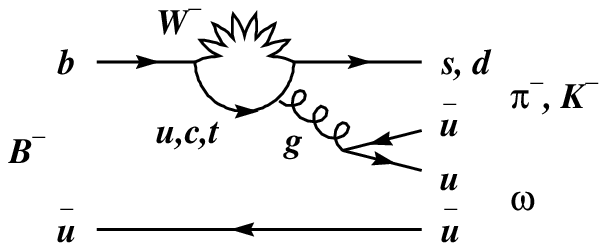, width=1.65in}
\end{center}
\caption{Tree (left) and penguin (right) diagrams for
$B^- \to \omega K^-$ and $B^-\to\omega \pi^-$ decays.}
\label{diagram}
\end{figure}

Charmless hadronic $B$ decays are of interest not only for testing
our current understanding of heavy quark physics, but also as
modes to search for direct $CP$ violation. The $B^-
\rightarrow \omega \pi^-$ and $\omega K^-$ decays~\cite{ch-conj} are
dominated by tree-level and gluonic penguin diagrams~\cite{gfact},
respectively, illustrated in Fig.~\ref{diagram}. Thus, their
branching fractions can give us further insight into gluonic
penguin diagrams, while interference between tree and penguin
diagrams can lead to a measurable direct $CP$ asymmetry.

In factorization models with $N_c \simeq 2$--3, where $N_c$
is the effective number of colors, the
$\omega\pi$ mode is larger than the $\omega K$ mode by a factor of
2 or more~\cite{gfact}. This result is borne out by further studies
in the QCD factorization~\cite{Du:2002up} and perturbative QCD
(pQCD)~\cite{pQCD} frameworks.
In this letter, we report measurements of $B^- \to
\omega K^-$ and $\omega \pi^-$ decays that indicate that 
the former is more prominent, which may suggest the influence of
nonfactorized effects.

The $B^- \to \omega K^-$ mode has an interesting history. It was
first reported by CLEO in 1998~\cite{cleo-prl-81} with 3.9$\sigma$
significance, but subsequently superseded by non-observation with
a larger data set~\cite{cleo-prl-85}, a result that is supported
by {\sc BaBar}~\cite{babar-prl-87}. However, we report here a
significant signal in this mode. The $B^-\to \omega \pi^-$ mode
has been reported previously by the CLEO~\cite{cleo-prl-85} and
{\sc BaBar}~\cite{babar-prl-87} collaborations at levels that are
somewhat higher than our findings. The data used in this analysis
were collected with the Belle detector~\cite{NIM} at
KEKB~\cite{kekb}, a double storage ring that collides 8 GeV
electrons and 3.5 GeV positrons with a 22 mrad crossing angle. The
data sample corresponds to an integrated luminosity of $29.4$
fb$^{-1}$ on the $\Upsilon(4S)$ resonance, containing 31.9 million
$B\overline{B}$ pairs, and 2.3 fb$^{-1}$ taken 60 MeV below the
resonance.

Belle is a general-purpose detector with a 1.5 T superconducting
solenoid magnet.
Charged particle tracking, covering 86\% of the total center-of-mass (CM)
solid angle, is provided by a Silicon Vertex Detector (SVD) consisting of
three concentric layers of double-sided silicon strip detectors,
and a 50-layer Central Drift Chamber (CDC).
Charged hadrons are distinguished by combining the responses
from an array of Silica Aerogel \v Cerenkov Counters (ACC),
a Time of Flight Counter system (TOF), and $dE/dx$ measurements in the
CDC.
The combined response provides $K/\pi$ separation of at least 2.5$\,\sigma$
for laboratory momenta up to 3.5 GeV/$c$.
Photons and electrons are detected in an array of 8736
CsI(Tl)
crystals (ECL) located inside the magnetic field and
covering the entire solid angle of the charged particle tracking system.
The 1.5~T magnetic field is returned via a flux return that consists of
4.7 cm thick steel plates interspersed with resistive plate
chambers to detect muons and $K_L$ mesons (KLM).
The Belle detector is described in detail elsewhere~\cite{NIM}.

Well reconstructed tracks that are inconsistent with being electrons or
muons are identified as kaon or pions according to a
$K/\pi$ likelihood ratio (KID),
${\cal L}_K / ({\cal L}_{\pi} + {\cal L}_K)$, where the
${\cal L}_{K(\pi)}$ are likelihoods derived
from the responses of the $dE/dx$, ACC and TOF systems.
Candidate $\pi^0$ mesons are reconstructed from pairs
of photons, each consisting of energy clusters greater than
50 MeV in the ECL, with $m_{\gamma\gamma}$
inside a $\pm 3\sigma$ ($\sigma = 5.4~{\rm MeV}/c^2$)
mass window around the $\pi^0$ mass~\cite{pdg}.
A mass-constrained fit is then performed to improve the
$\pi^0$ momentum resolution.
Candidate $\omega$ mesons are formed from $\pi^+\pi^-\pi^0$
combinations with an invariant mass that is within $\pm 30$ MeV/$c^2$ of the nominal $\omega$ mass~\cite{pdg}.
(The natural width of the $\omega$ meson is $8.9$ MeV.)
To futher reduce the large combinatorial background from
low energy photons and $\pi^0$s, an $\omega$ candidate is discarded if
the daughter $\pi^0$'s CM momenta is below 350 MeV/$c$.
This selection on $\pi^0$ CM momentum loses 16\% of the signal,
but removes 60\% of the combinatorial background.



We combine an $\omega$ candidate with either a $K^-$ or a $\pi^-$ track to
form a ${B}^-$ candidate.
As part of this procedure, the momenta of the
three charged tracks are recalculated subject to the constraint that they
originate from the interaction point.
Using the CM beam energy
$E^{\rm CM}_{\rm beam} = \sqrt{s}/2 = 5.29$ GeV and the measured CM energy
$E^{\rm CM}_B$ and momentum $p^{\rm CM}_B$ of the $B$ candidate, we
form two kinematic variables to select the signal events:
the beam-constrained mass
$M_{\rm bc} = \sqrt{(E_{\rm beam}^{\rm CM})^2 - (p_{B}^{\rm CM})^2}$
and the energy difference
$\Delta E = E_{B}^{\rm CM} - E_{\rm beam}^{\rm CM}$.


The major background for this analysis is from continuum
$e^+ e^- \to q\overline{q}$ production, where $q$ is a light quark
($u$, $d$, $s$, or $c$).
The jet-like continuum events are suppressed relative to the more spherical
$B\overline{B}$ events by characterization of the event shape,
which is implemented with a Fisher discriminant~\cite{fisher} containing
six modified Fox-Wolfram moments~\cite{hh,fw}.
There are two types of combinatorial backgrounds from continuum events:
fake $\omega$ mesons and fake $B$ mesons.
The former is suppressed using the
cross product $|\vec{P}_{+} \times \vec{P}_{-}|$ of the momenta of
the charged pion daughters in the $\omega$ meson rest frame.
The latter is suppressed using the $B$ candidate flight direction
relative to the positron beam axis, and the helicity angle
of the candidate $\omega$ meson relative to the $B$ meson.
The helicity angle, $\theta_{\rm hel}$,
is defined as the angle between the $B$ flight
direction and the vector perpendicular to the $\omega$ decay plane in the
$\omega$ rest frame.
We use a likelihood ratio technique that combines the
Fisher discriminant, the cross product of the momenta of the charged pions
from the $\omega$, the $B$ flight direction, and the cosine of the
helicity angle, 
$\cos{\theta_{\rm hel}}$, to suppress
the continuum background relative to the
$B \to \omega h$ ($h = \pi$ or $K$) signal.
The probability density functions (PDFs) for signal and
background are constructed using Monte-Carlo (MC) events.
The background PDFs are in
good agreement with those determined from on-resonant sideband data
($M_{\rm bc} < 5.26$ GeV/$c^2$ and $|\Delta E| < 0.3$ GeV).
With these PDFs, we determine signal (${\cal L}_{\rm S}$)
and background (${\cal L}_{\rm BG}$) likelihoods
for each event that are used to form the normalized likelihood ratio
${\cal R} = {\cal L}_{\rm S} / ({\cal L}_{\rm S} + {\cal L}_{\rm BG})$;
we discard events with ${\cal R} < 0.85$.
This selection retains 50\% of the signal while rejecting 95\% of
the continuum background.


To study background from $B$ decays through the $b\to c$ transition and charmless $B$ decays such as $B \rightarrow \omega K^*$ and
$B \rightarrow \omega \rho$, and
non-resonant $B \rightarrow K^-\pi^+\pi^-\pi^0$ decays,
we used MC samples up to 20 times larger than our data sample,
assuming the best known branching fraction for each decay~\cite{cleo-vv}.
We find negligible backgrounds from these decays in the
$M_{\rm bc}$--$\Delta E$ signal region
($M_{\rm bc} > 5.27$ GeV/$c^2$ and $|\Delta E| < 0.1$ GeV).


The signal is extracted using $M_{\rm bc}$ and $\Delta E$ as independent
variables in an unbinned maximum likelihood fit for events with
$|\Delta E| < 0.3$ GeV and $M_{\rm bc} > 5.2$ GeV$/c^2$.
The signal PDF for $M_{\rm bc}$ is a Gaussian and that
for $\Delta E$ is the parameterization of Ref.~\cite{cbline}.
The parameters are determined from MC simulation and calibrated
by the decay chain $B^- \to D^0\pi^-$, $D^0 \to K^-\pi^+\pi^0$.
The resolutions determined from MC are 3 MeV/$c^2$ for
$M_{\rm bc}$ and 24 MeV for $\Delta E$.
The PDF of continuum background for $M_{\rm bc}$ is
an empirically determined threshold function~\cite{argus}
that is obtained from the sideband data ($\Delta E > 0.1$ GeV),
while the PDF for $\Delta E$ is a linear polynomial
obtained from the data ($M_{\rm bc} > 5.27$ GeV$/c^2$) before the $\cal R$ cut.
PDFs for other background sources are included:
charmless $B$ decays that survive the selection criteria and
signal events with charged kaons
misidentified as pions or vice versa.
(In the last case, the PDFs have the same shape as signal except that
the central value of $\Delta E$ is shifted by 45 MeV.)

\begin{figure}[t]
\begin{center}
\epsfig{file=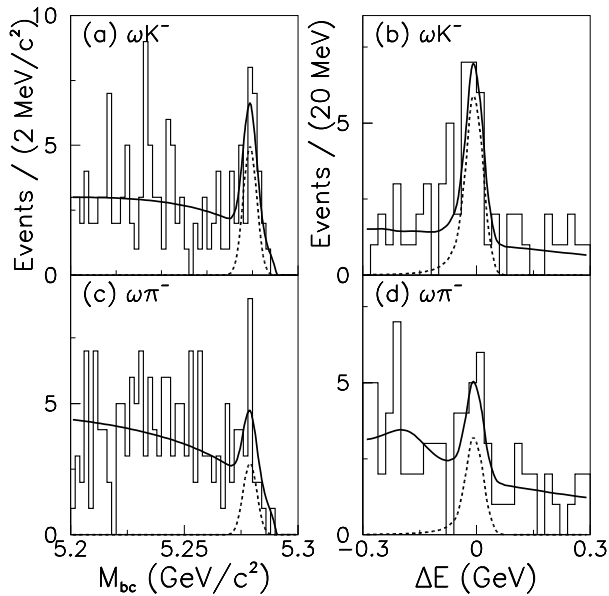, width=3.4in}
\end{center}
\caption{The $M_{\rm bc}$ (left) and $\Delta E$ (right) distributions
of the candidate events (histograms), the best fits (solid curves)
and signal components (dashed curves).}
\label{mbde}
\end{figure}

The signal yields from the maximum likelihood fit are
summarized in Table~\ref{yields-table}.
The $M_{\rm bc}$ and $\Delta E$ distributions of candidate events and
the best fit curves are shown in Fig.~\ref{mbde}.
The signal yields are $18.9^{+5.4}_{-4.7}$ and $10.4^{+4.7}_{-4.3}$
events for the $\omega K^-$ and $\omega\pi^-$ modes, respectively.
The expected reflection due to $\pi$-$K$ misidentification
is $0.7\pm 0.3$ ($2.0\pm0.6$) events
for the $\omega K^- (\omega \pi^-)$ mode; the fit
gives $0.0\pm 2.4$ ($0.0\pm 3.9$) events.

\begin{table}[t]
\caption{ The signal yields, statistical significances ($\Sigma$),
 efficiencies ($\epsilon$), branching fractions ($\cal{B}$),
and the $90\%$ confidence level upper limit (UL) of the branching
fraction for the $\omega\pi^-$ mode are listed.
The efficiencies include the $\omega$ decay branching fraction.
}
\medskip
\begin{tabular}{lccccc} \hline\hline
    & Signal yield & $\Sigma$
    &  $\epsilon$ (\%)
    & $\cal B$ {\small($\times 10^{-6}$)}
    & UL {\small($\times 10^{-6}$)} \\ \hline
$\omega K^-$    & $18.9^{+5.4}_{-4.7}\pm 0.6$ & 6.0$\,\sigma$
        & $6.0$
        & $9.2^{+2.6}_{-2.3}\pm 1.0$ & -\\
$\omega \pi^-$  & $10.4^{+4.7}_{-4.3}{}^{+0.4}_{-0.6}$ & 3.3$\,\sigma$
        & $7.7$
        & $4.2^{+2.0}_{-1.8}\pm 0.5$
        & 8.1\\ \hline\hline
\end{tabular}
\label{yields-table}
\vskip1pc
\end{table}


The statistical significance quoted in Table~\ref{yields-table}
is defined as $\sqrt{-2{\rm ln}({\cal L}(0)/{\cal L}_{\rm max})}$
where ${\cal L}_{\rm max}$ is the maximized likelihood at the nominal signal
yield and ${\cal L}(0)$ is the likelihood with the signal yield fixed
at zero.  We observe 18.9 signal events for $B^- \to \omega K^-$
with $6.0\,\sigma$ significance and find 10.4 $\omega \pi^-$ events
with $3.3\,\sigma$ significance. Since the latter has less than
$4\,\sigma$ significance,
we use the 90\% confidence level upper limit ($N_S^{UL}$) of
17.3 events on $B^- \to \omega \pi^-$ yield, determined by integrating
the likelihood as a function of the number of signal events
to 90\% of its total area.

We study the systematic error associated with the fit by
varying the parameters in the fitting functions by $1\sigma$
from their nominal values. The change in the signal yield from each
variation is added in quadrature to obtain an overall systematic error
associated with the fit.
The systematic errors in the detection efficiencies
of the $\omega$ meson and the high-momentum $K^-$ and $\pi^-$ mesons are
8.5\% and 2.2\%, respectively, which are
 determined from detailed studies of charged particle tracking,
$\pi^0$ detection, and particle identification.
A 5\% systematic uncertainty is assigned to the continuum
suppression cut, which is obtained by applying a similar procedure to data
and MC samples of $B^- \rightarrow D^{*0}\pi^-$ events.
The combined uncertainty of the efficiency is 10.1\%.

The branching fractions in Table~\ref{yields-table} are calculated
assuming equal numbers of $B^+B^-$ and $B^0\overline{B}{}^0$ pairs
in our data sample.
The uncertainty in the number of $B\overline{B}$ events, 1\%, is taken into
account and included in the systematic error for the branching fraction.
The upper limit of the branching fraction of
$\omega\pi^-$ decay is calculated after increasing $N_S^{UL}$ and
reducing the efficiency by their respective systematic error.

\begin{figure}[t]
\begin{center}
\epsfig{file=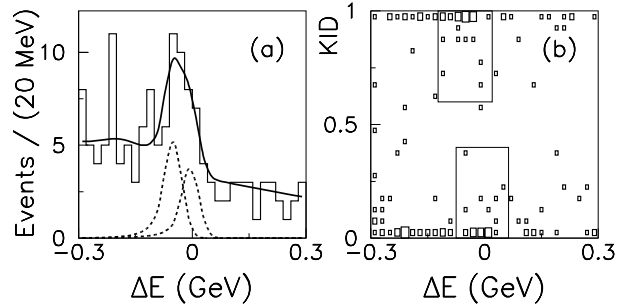, width=3.4in}
\end{center}
\caption{The (a) $\Delta E$ distribution and (b) scatter plot of
KID likelihood ratio versus $\Delta E$ for the $\omega h^-$ mode.
The solid curve
shows the fit result with the signal components shown by dashed curves.}
\label{de-nokid}
\end{figure}

Our branching fraction result for $B^-\to \omega K^-$ is
larger than that for $B^-\to \omega\pi^-$. As a consistency check,
we also performed the analysis without KID information.
Figure~\ref{de-nokid} shows
the $\Delta E$ distribution and a scatter plot of the KID likelihood ratio
versus $\Delta E$ for the $\omega h^-$ candidates. In these plots, we use
the $\pi^-$ mass for the high momentum hadron track. This causes a
$-45$~MeV difference between the peak positions of $\omega K^-$ and
$\omega\pi^-$ signals.
The $\Delta E$ distribution is fitted with $\omega K^-$ and $\omega
\pi^-$ signals, continuum background, and charmless background
components.
The signal yields are $17.1 \pm 7.7$ and $12.1 \pm 7.0$ events for $\omega
K^-$ and $\omega\pi^-$, respectively, and are consistent with the
results using the KID for $K/\pi$ separation.
The scatter plot in Fig.~\ref{de-nokid}(b)
shows the distribution of events in KID versus $\Delta E$.
The large rectangles, which cover the $\pm3 \sigma$ signal regions in $\Delta E$ and high or low
kaon probability, contain enhancements at the appropriate places for both modes.

\begin{figure}[t]
\begin{center}
\epsfig{file=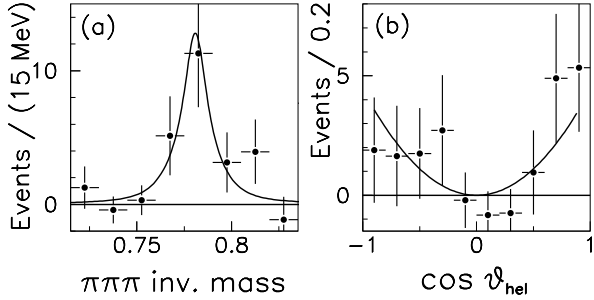, width=3.4in}
\end{center}
\caption{The $B^-\to \omega K^-$ signal yield in bins of
(a) $\pi^+\pi^-\pi^0$ invariant mass and
(b) cosine of the $\omega$ helicity angle.
The solid curve shows the fit result.}
\label{omega}
\end{figure}

We also examine the properties of the $\omega$ candidates to confirm
the $B^- \to \omega K^-$ signal.
The $B^-\to \omega K^-$ signal yield in
$\pi^+\pi^-\pi^0$ invariant mass bins
is shown in Fig.~\ref{omega}(a).
A clear signal at the $\omega$ mass is seen.
The fitted number of $\omega$ mesons is $18.0 \pm 5.0$
which is consistent with the $\omega K^-$ signal yield.
Figure~\ref{omega}(b) shows the $B^-\to\omega K^-$ signal yield in
$\cos{\theta_{\rm hel}}$ bins. The requirement on
the likelihood ratio has been applied without including the helicity
angle variable. The distribution is consistent with the expected
$\cos^2\theta_{\rm hel}$ distribution.

We also measure the partial rate asymmetry in $B^{\pm}\to\omega K^{\pm}$
decays to search for direct $CP$ violation.
The asymmetry is defined as
\[ {\cal A}_{CP}
= \frac{N(\omega K^-) - N(\omega K^+)}{N(\omega K^-) + N(\omega K^+)}.
\]
An application of the same event extraction and fitting
procedure to the $B^-$ and $B^+$ candidates separately
yields $7.3\pm 3.5$ and $11.2 \pm 3.7$ events for
$\omega K^-$ and $\omega K^+$, respectively, and an
asymmetry value ${\mathcal A}_{CP} = -0.21 \pm 0.28 \pm 0.03$.
The systematic error includes the uncertainty associated with the fit procedure as well as a contribution of 1\% due to detector bias in reconstruction of positive and negative high-momentum kaon tracks.
The 90\% confidence level interval $-0.70 < {\cal A}_{CP} < 0.28$
is obtained by assuming a Gaussian statistical error convolved with the systematic error.

Our combined branching fraction of
$(13.4^{+3.3}_{-2.9}\pm 1.1) \times 10^{-6}$ for
$B^- \to \omega h^-$ ($h = \pi$ or $K$) agrees with CLEO's number,
$(14.3^{+3.6}_{-3.2}\pm 2.0) \times 10^{-6}$~\cite{cleo-prl-85},
although the individual branching fractions are not totally consistent.
Our large $B^- \to \omega K^-$ branching fraction
also disagrees with the upper limit of $4\times 10^{-6}$ reported by
the {\sc BaBar} collaboration~\cite{babar-prl-87},
although our $B^-\to \omega\pi^-$ result is not in conflict.
We note that {\sc BaBar}'s combined branching fraction for
$B^- \to \omega h^-$ ($h = \pi$ or $K$)
is low compared to CLEO and our result.

The large $B^- \to \omega K^-$ branching fraction and 
relatively low $B^- \to \omega\pi^-$ rate 
cannot be easily accounted for either by generalized
factorization~\cite{gfact} with $N_c \simeq 2$--3 or by
calculations based on pQCD~\cite{Du:2002up,pQCD}.
To accommodate the large $B^- \to \omega K^-$ branching fraction that we
observe, it appears that $N_c$ has to deviate significantly from
3~\cite{nonfact}, indicating the presence of large non-factorizable
effects. 

In summary, using 31.9 million $B\overline{B}$ pairs collected
with the Belle detector, we report the first observation of
the $B^-\to\omega K^-$ decay with branching fraction
${\cal B} ( B^- \to \omega K^- ) =
    (9.2{}^{+2.6}_{-2.3}\pm 1.0) \times 10^{-6}$;
the statistical significance of the above signal is $6.0\sigma$.
We also measure ${\cal B} ( B^- \to \omega \pi^- ) =
    (4.2{}^{+2.0}_{-1.8}\pm 0.5) \times 10^{-6}$,
with a statistical significance of $3.3\sigma$.
The partial rate asymmetry for $B^{\pm} \to \omega K^{\pm}$ decays
is found to be ${\cal A}_{CP} = -0.21 \pm 0.28 \pm 0.03$, corresponding to
a 90\% confidence level interval of $-0.70 < {\cal A}_{CP} < 0.28$.

We wish to thank the KEKB accelerator group for the excellent
operation of the KEKB accelerator.
We acknowledge support from the Ministry of Education,
Culture, Sports, Science, and Technology of Japan
and the Japan Society for the Promotion of Science;
the Australian Research Council
and the Australian Department of Industry, Science and Resources;
the National Science Foundation of China under contract No.~10175071;
the Department of Science and Technology of India;
the BK21 program of the Ministry of Education of Korea
and the CHEP SRC program of the Korea Science and Engineering Foundation;
the Polish State Committee for Scientific Research
under contract No.~2P03B 17017;
the Ministry of Science and Technology of the Russian Federation;
the Ministry of Education, Science and Sport of the Republic of Slovenia;
the National Science Council and the Ministry of Education of Taiwan;
and the U.S.\ Department of Energy.


\end{document}

%% file: author.tex
\affiliation{Aomori University, Aomori}
\affiliation{Budker Institute of Nuclear Physics, Novosibirsk}
\affiliation{Chiba University, Chiba}
\affiliation{Chuo University, Tokyo}
\affiliation{University of Cincinnati, Cincinnati OH}
\affiliation{Gyeongsang National University, Chinju}
\affiliation{University of Hawaii, Honolulu HI}
\affiliation{High Energy Accelerator Research Organization (KEK), Tsukuba}
\affiliation{Hiroshima Institute of Technology, Hiroshima}
\affiliation{Institute of High Energy Physics, Chinese Academy of Sciences, Beijing}
\affiliation{Institute of High Energy Physics, Vienna}
\affiliation{Institute for Theoretical and Experimental Physics, Moscow}
\affiliation{J. Stefan Institute, Ljubljana}
\affiliation{Kanagawa University, Yokohama}
\affiliation{Korea University, Seoul}
\affiliation{Kyoto University, Kyoto}
\affiliation{Kyungpook National University, Taegu}
\affiliation{Institut de Physique des Hautes \'Energies, Universit\'e de Lausanne, Lausanne}
\affiliation{University of Ljubljana, Ljubljana}
\affiliation{University of Maribor, Maribor}
\affiliation{University of Melbourne, Victoria}
\affiliation{Nagoya University, Nagoya}
\affiliation{Nara Women's University, Nara}
\affiliation{National Kaohsiung Normal University, Kaohsiung}
\affiliation{National Lien-Ho Institute of Technology, Miao Li}
\affiliation{National Taiwan University, Taipei}
\affiliation{H. Niewodniczanski Institute of Nuclear Physics, Krakow}
\affiliation{Nihon Dental College, Niigata}
\affiliation{Niigata University, Niigata}
\affiliation{Osaka City University, Osaka}
\affiliation{Osaka University, Osaka}
\affiliation{Panjab University, Chandigarh}
\affiliation{Peking University, Beijing}
\affiliation{Saga University, Saga}
\affiliation{University of Science and Technology of China, Hefei}
\affiliation{Seoul National University, Seoul}
\affiliation{Sungkyunkwan University, Suwon}
\affiliation{University of Sydney, Sydney NSW}
\affiliation{Tata Institute of Fundamental Research, Bombay}
\affiliation{Toho University, Funabashi}
\affiliation{Tohoku Gakuin University, Tagajo}
\affiliation{Tohoku University, Sendai}
\affiliation{University of Tokyo, Tokyo}
\affiliation{Tokyo Institute of Technology, Tokyo}
\affiliation{Tokyo Metropolitan University, Tokyo}
\affiliation{Tokyo University of Agriculture and Technology, Tokyo}
\affiliation{Toyama National College of Maritime Technology, Toyama}
\affiliation{University of Tsukuba, Tsukuba}
\affiliation{Utkal University, Bhubaneswer}
\affiliation{Virginia Polytechnic Institute and State University, Blacksburg VA}
\affiliation{Yokkaichi University, Yokkaichi}
\affiliation{Yonsei University, Seoul}
  \author{R.-S.~Lu}\affiliation{National Taiwan University, Taipei} 
  \author{K.~Abe}\affiliation{High Energy Accelerator Research Organization (KEK), Tsukuba} 
  \author{K.~Abe}\affiliation{Tohoku Gakuin University, Tagajo} 
  \author{N.~Abe}\affiliation{Tokyo Institute of Technology, Tokyo} 
  \author{R.~Abe}\affiliation{Niigata University, Niigata} 
  \author{T.~Abe}\affiliation{Tohoku University, Sendai} 
  \author{I.~Adachi}\affiliation{High Energy Accelerator Research Organization (KEK), Tsukuba} 
  \author{H.~Aihara}\affiliation{University of Tokyo, Tokyo} 
  \author{Y.~Asano}\affiliation{University of Tsukuba, Tsukuba} 
  \author{T.~Aso}\affiliation{Toyama National College of Maritime Technology, Toyama} 
% ...
\collaboration{The Belle Collaboration}

%% file: omegah.bbl
\begin{thebibliography}{99}
\bibitem{ch-conj}{
    Charge conjugate modes are implicitly included throughout the
    paper.
}
\end{thebibliography}
